\documentclass[runningheads]{llncs}
\usepackage[T1]{fontenc}
\usepackage{graphicx}

\usepackage{array} 

\usepackage{tabularx} 

\usepackage{longtable}

\usepackage{hyperref}
\begin{document}
\title{Self-Service or Not? How to Guide Practitioners in Classifying AI Systems Under the EU AI Act}
\titlerunning{Self-Service or Not?}
%
\author{Ronald Schnitzer\inst{1}\inst{2} \and Maximilian Hoeving \inst{2}\and Sonja Zillner\inst{1}\inst{2}}
\institute{Technical University of Munich, School of Computation, Information and Technology, Munich, Germany \\
\email{ronald.schnitzer@tum.de}
\and
Siemens AG, Munich, Germany \\
\email{maximilian.hoeving@siemens.com}\\
\email{sonja.zillner@siemens.com}
}

\maketitle              

\begin{abstract}
In August 2024, the EU Artificial Intelligence Act (AIA) came into force, marking the world’s first large-scale regulatory framework for AI. Central to the AIA is a risk-based approach, aligning regulatory obligations with the potential harm posed by AI systems. To operationalize this, the AIA defines a Risk Classification Scheme (RCS), categorizing systems into four levels of risk. While this aligns with the theoretical foundations of risk-based regulations, the practical application of the RCS is complex and requires expertise across legal, technical, and domain-specific areas. Despite increasing academic discussion, little empirical research has explored how practitioners apply the RCS in real-world contexts.
This study addresses this gap by evaluating how industrial practitioners apply the RCS using a self-service, web-based decision-support tool. Following a Design Science Research (DSR) approach, two evaluation phases involving 78 practitioners across diverse domains were conducted. Our findings highlight critical challenges in interpreting legal definitions and regulatory scope, and show that targeted support, such as clear explanations and practical examples, can significantly enhance the risk classification process. The study provides actionable insights for tool designers and policymakers aiming to support AIA compliance in practice.
\textbf{Note:} This is the submitted version of the manuscript. 
The paper has been accepted and presented at the 3rd International Conference on Frontiers of Artificial Intelligence, Ethics, and Multidisciplinary Applications (FAIEMA), 2025. The final Version of Record will appear in the conference proceedings.
\end{abstract}
\section{Introduction}
\label{sec: Introduction}

With the rapid advancement of Artificial Intelligence (AI), new risks to individuals and society have emerged. In response, the European Union developed harmonized rules to govern the development and deployment of AI systems. After years of political negotiation and legislative drafting, the European AI Act (AIA) was formally enacted in August 2024 \cite{AI_Act}.

A central principle of the AIA is its risk-based regulatory approach. Rather than enforcing rigid rules, this framework seeks to manage risks proportionately to the potential harm posed by a given AI system \cite{black_really_2010}. By doing so, the EU aims to ensure the safe and trustworthy use of AI while continuing to support innovation. Accordingly, the AIA aligns the level of regulatory obligation with the assessed risk level of an AI system.

Specifically, the AIA defines four risk levels: Unacceptable Risk (e.g., social scoring), High-Risk (e.g., safety components in critical infrastructure), Limited-Risk (e.g., chatbots), and Low-Risk (e.g., non-interactive classifiers used in non-sensitive domains).

Accurate classification of AI systems under the AI Act is essential for practitioners, as compliance with the regulation is legally mandated and non-compliance may result in significant penalties (AIA Art. 99).
While the classification rules are primarily defined in Articles 5 (Prohibited AI practices), 6 (High-risk AI systems), and 50 (Transparency obligations), applying them in practice is far from straightforward. The classification logic—referred to in this study as the AIA \textbf{Risk Classification Scheme (RCS)}—has already been shown in theoretical analyses to be complex and to require specialized expertise. Moreover, prior literature highlights several logical ambiguities and inconsistencies within the RCS \cite{hupont_landscape_2022,leinarte_classification_2024,martini_ki-vo_2024,neuwirth_prohibited_2023}, further complicating its application.
Despite these challenges, no empirical studies have yet examined how the RCS is applied in practice, particularly by practitioners without legal or regulatory expertise.

To close this gap, this study aims to derive insights and enhance the application of the RCS in an industrial setting by addressing the following research questions: 
\begin{description}
   \item[RQ1] What level of expertise is required to effectively apply the RCS? \item[RQ2] What are the key challenges practitioners face in the RCS, and how can these be addressed?
\item[RQ3] What specific information or resources would enhance practitioners' ability to apply the RCS effectively?
\end{description}

In addressing these research questions, this study contributes by:

\begin{description}
\item[(i)] collecting empirical data on practitioners and their behavior when applying the RCS in real-world settings,
\item[(ii)] identifying key challenges faced by practitioners and highlighting areas where legislators should provide additional clarification and guidance, and
\item[(iii)] deriving actionable design knowledge for the implementation of the RCS in self-service tools.

\end{description}

This research involves the development and utilization of a web-based decision-support tool implementing the RCS, followed by a systematic analysis of practitioners' application of the tool to derive qualitative and quantitative insights into the regulation's risk classification process.

The remainder of this work is structured as follows: Section \ref{sec: Related Work} positions this work within the existing body of research related to the AIA and its RCS.
Section \ref{sec: Methodology} provides a detailed account of the research methodology employed.
Section \ref{sec: Evaluation Results} presents and discusses the study's results, followed by an examination of the study's limitations and implications in Section \ref{sec: Discussion and Limitations}.
The paper concludes with final remarks in Section \ref{sec: Conclusion}.

\section{Related Work}
\label{sec: Related Work}
The related work relevant to this study can be grouped into four categories: (1) literature on risk-based regulation, (2) general scholarship on the AI Act and specific aspects of the regulation, including its risk classification scheme, (3) research on translating legal texts and requirements into software, and (4) existing tools designed to support and studies analyzing the implementation of the RCS. The remainder of this section provides an overview of each of these areas.

\paragraph{Risk-based regulation}

While this study centers on the RCS in the AIA, broader work on risk-based regulation provides valuable context. For instance, Rothstein et al. \cite{rothstein_risks_2006} identify several structural issues inherent in risk-based regulatory approaches.

Unlike regulations that rely on qualitative thresholds to determine regulatory obligations, such as the GDPR, the AIA defines distinct risk categories that directly determine the applicable legal requirements. A comparable approach is found in the EU Medical Device Regulation (MDR), which categorizes devices into multiple risk classes and provides detailed, rule-based criteria for classification. However, in contrast to the MDR’s relatively well-established framework, the AIA introduces greater interpretive uncertainty. This may be attributed to the fact that MDR builds upon decades of regulatory practice and continuity from the Medical Devices Directive (MDD), whereas the AIA addresses a rapidly evolving technological domain with limited prior regulatory precedent.

\paragraph{The European AI Act}

While the final version of the AIA was only published in July 2024 \cite{AI_Act}, much of the existing scholarship is based on earlier drafts. While some aspects may have evolved, we consider research related to earlier versions still valuable and thus include it in this section.

Several works provide overviews of the regulation and analyze its potential impacts. For instance, Veale et al. \cite{veale_demystifying_2021} offer a comprehensive overview of the first draft, while Ebers \cite{ebers_european_2021} presents a critical examination of its prospective effects. The overview is complemented by Martini et al. \cite{martini_ki-vo_2024}, a comprehensive legal commentary on the AIA.

Further investigations focus on specific sections of the regulation. Wendehorst et al. \cite{wendehorst_begriff_2024} analyze the definition of AI systems and assess their individual components from a legal perspective. Schuett \cite{schuett_risk_2024} examines Article 9 on Risk Management, whereas Hacker et al. \cite{hacker_regulating_2023} analyze the regulatory challenges posed by technological advances, such as Generative AI and foundation models. Sovrano \cite{sovrano_metrics_2022} evaluates the applicability of Explainable AI methods and metrics for compliance with the regulation.

The RCS has been a subject of extensive academic discourse. Barkane \cite{barkane_questioning_2022} criticizes the RCS, suggesting that Emotion Recognition Systems require stricter regulation. Other scholars highlight challenges and inconsistencies within the scheme. Neuwirth \cite{neuwirth_prohibited_2023} critically discusses the proposed regulatory mechanics behind the unacceptable risk category. Hupont \cite{hupont_landscape_2022} analyzes a wide array of AI systems in biometric identification, showing that not all use cases can be unambiguously assigned to one of the categories, hence pointing out the non-exclusive nature of the risk categories. Given the rules laid down by the AIA, Martini et al. \cite{martini_ki-vo_2024} point out that AI systems can be simultaneously classified as high- and limited-risk. Leinarte \cite{leinarte_classification_2024} identifies, as part of a broader analysis, inconsistencies in how central terms are defined in the AIA compared to other Union legislation.

\paragraph{Software Tooling for Legal Applications}

Another key aspect of this study is the translation of legal terminology and regulatory requirements into a functional software tool. Escher et al. \cite{escher_code-ifying_2024} present a case study in which sections of the U.S. Bankruptcy Code are operationalized in software, emphasizing the importance of multidisciplinary collaboration between legal and technical experts to ensure accurate and faithful implementation. Furthermore, the human factor plays a critical role in the effective use of such tools.
Cranor \cite{cranor_framework_2008} highlights the challenges of human misunderstanding in socio-technical systems, particularly in the context of security-related applications. These insights support the view that regulatory software must be both legally sound and usable for non-expert practitioners.

\paragraph{Implementations of the RCS}

Building on these foundations, several tools implementing the RCS have recently emerged, such as the EU AI Act Compliance Checker \cite{future_of_life_institute_eu_2025} and the TUEV AI Risk Navigator \cite{tuev_ai_lab_ai_2025}. This highlights the demand for practical support mechanisms to implement the RCS. However, to the best of our knowledge, no systematic analysis of the effectiveness of these tools has been published yet.

The study most closely related to ours is by Hanif et al. \cite{hanif_navigating_2024}, who developed a decision tree representation of the RCS to evaluate whether non-legal experts could accurately classify AI systems. Their work provided early evidence of the feasibility of self-service classification, but it was based on a pre-final draft of the AIA and focused primarily on classification accuracy. In contrast, our study uses the final version of the regulation and shifts the focus toward understanding the conditions under which practitioners can successfully apply the RCS. Specifically, we investigate the types of expertise required, identify usability and comprehension challenges, and evaluate which forms of support (e.g., examples, definitions, expert access) improve classification confidence. As such, this study addresses both conceptual ambiguities in the RCS and practical challenges in its industrial application.

\section{Methodology}
\label{sec: Methodology}

\begin{figure}
    \centering
    \includegraphics[width=\linewidth]{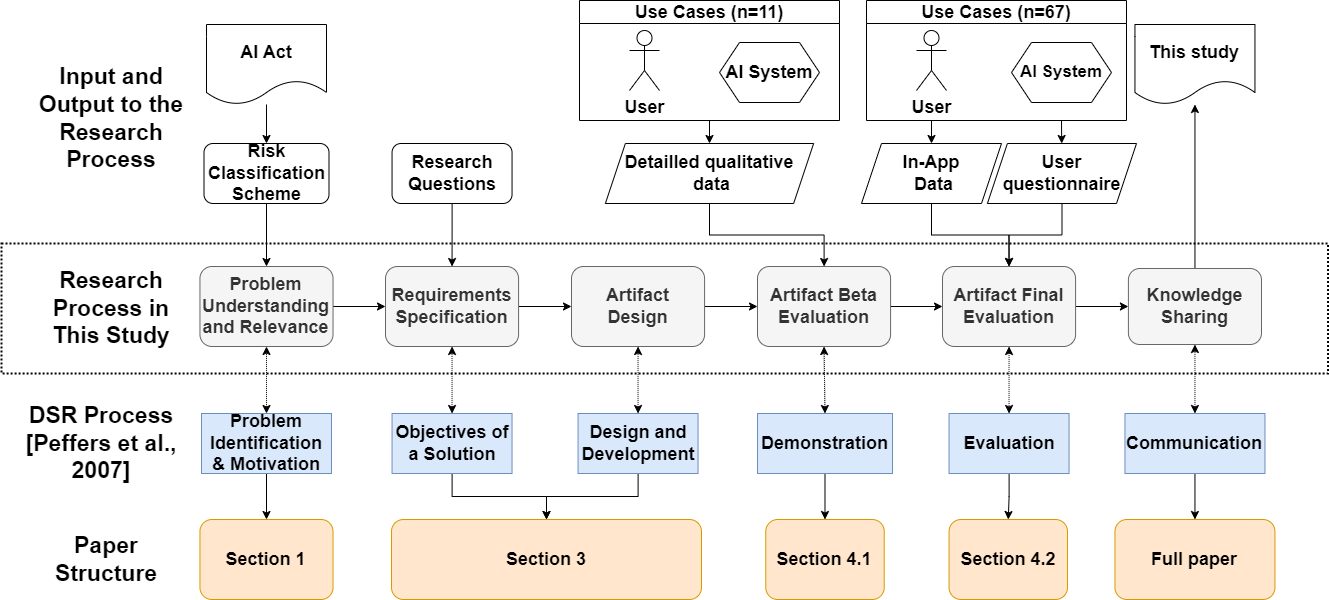}
    \caption{Research process of this study with inputs and outputs shown at the top.  Below, each step is mapped to the DSR process \cite{peffers_design_2007}, and the related section in this paper.}
    \label{fig: Research Process}
\vspace{-2ex}
\end{figure}

This study adopts a mixed-methods approach grounded in Design Science Research (DSR), which is well-suited for addressing real-world problems through the design and evaluation of an artifact \cite{henver_design_2004,peffers_design_2007}. Following the DSR model by Peffers et al.~\cite{peffers_design_2007}, we developed and evaluated a software artifact to support the application of the RCS. Figure~\ref{fig: Research Process} outlines the process. 

The artifact was designed and evaluated by a multidisciplinary team including legal, regulatory, and AI experts. Evaluation combined qualitative and quantitative data across two distinct study phases, as detailed in Section~\ref{subsec: Evaluation Design}.

\subsection{Artifact Design}
\label{subsec: Artifact Design}
To align with the study’s objectives, the web-based tool was designed to fulfill two main requirements:
\begin{itemize}
    \item[\textbf{(i)}] Practical utility: Accurately implement the classification logic defined in the AIA and support practitioners in independently applying the RCS.
    \item[\textbf{(ii)}] Research validity: Enable systematic evaluation by capturing reproducible data on user interactions and classification decisions.
\end{itemize}

To meet requirement (i), we implemented a decision tree-based approach, similar to the structure proposed by Hanif et al.~\cite{hanif_navigating_2024}. The classification logic is represented as a structured sequence of questions with predefined answers, guiding users toward the appropriate risk category based on their selections.

We derived the decision logic by qualitatively analyzing the final text of the AIA \cite{AI_Act}.
Following Articles 5, 6, and 50 of the AIA, we derived a four-category structure (unacceptable, high, limited, and low risk) consistent with prior literature \cite{hanif_navigating_2024}.

Our analysis identified two important additions to the RCS. First, we introduced an “out of scope” category for systems that either do not meet the AIA’s AI definition (Art. 3.1) or fall under explicit exclusions such as military or scientific use (Art. 2). Second, consistent with observations in prior literature \cite{martini_ki-vo_2024}, we allowed for non-exclusive classification, as systems may simultaneously meet the criteria for high-risk (Art. 6) and limited-risk (Art. 50).

\begin{figure}
    \centering
    \includegraphics[width=0.8\linewidth]{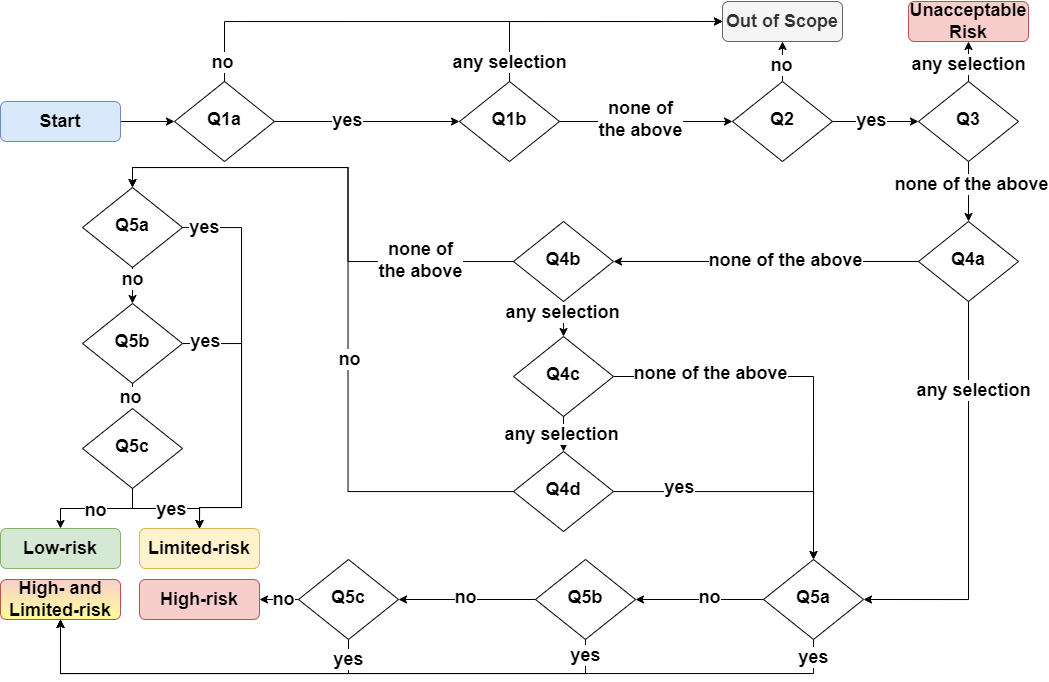}
    \caption{Decision-Tree representation of the implemented RCS.}
    \label{fig: Logic}
\end{figure}

Based on the identified categories, we developed a set of questions and answer options to represent the RCS decision logic. Questions were designed to be easily interpretable while preserving legal fidelity. Legal correctness was validated by two legal experts supporting this study. The complete decision logic is shown in Figure~\ref{fig: Logic} and the question-answer pairs are given in Table~\ref{tab: DT_final}.

\begin{table}
    \caption{Question-answer pairs representing the implemented RCS.
    \newline Answers: y/n = yes/no, MS = Multi-selection + option "none of the above"}
    \centering
    \begin{tabular}{|p{15 pt}|p{260 pt}|p{20 pt}|p{35 pt}|}
    \hline
        \textbf{ID} & \textbf{Question} &\textbf{Ans.}& \textbf{AIA Ref.}  \\
    \hline
            Q1a &Is your AI system covered by the geographical scope of the AI Act either: 
            
            (i) by being placed on the market or put into service within the EU; or
            (ii) by producing outputs that are used within the EU; or
            (iii) or by affecting individuals located in the EU?
             &y/n& Art. 2 \\
    \hline
            Q1b & Does one of the following conditions apply to your system?

            <Selectable list of exemption conditions as defined in Art. 2> &MS& Art. 2 \\

    \hline
            Q2 &Do you make use of an AI system, i.e., a machine-based system that:   
 (i) Generates outputs based on input data, and 
(ii) Makes decisions with some autonomy, and 
(iii) Infers responses rather than following only fixed rules &y/n& Art. 3.1 \\
    \hline
            Q3 &Is your AI system intended to... 
            
            <Selectable list of prohibited practices from Art. 5> & MS & Art. 5 \\
    \hline
            Q4a &Is your AI system...
            (i) ... itself required to undergo a third-party conformity assessment under the legislation listed below; or
            (ii) ...a safety component of a product that is required to undergo a third-party conformity assessment under the legislation listed below? 
            
            <Selectable list of harmonized legislation listed in Annex I>& MS& Art. 6.1; Annex I \\
    \hline
            Q4b &Will the AI system be part of the following use cases? If yes, select the use cases; if no, select none of the above.
            
            <Selectable list of AI systems listed in Annex III> & MS& Art. 6.2 ; Annex III \\
    \hline
            Q4c &Does one or more of the following statements apply?
            
            <Selectable list of conditions listed in Art. 6.3>&MS& Art. 6.3 \\
    \hline
            Q4d &Does your AI system perform profiling of natural persons? &y/n& Art. 6.3 \\
    \hline
            Q5a &Does your AI system directly interact with a natural person?&y/n& Art. 50.1 \\
    \hline
            Q5b &Could your AI system generate synthetic text, audio, image, video, or content? &y/n& Art. 50.2 \\
    \hline
            Q5c &Can your AI system be considered to be an emotion recognition or biometric categorisation system? &y/n& Art. 50.3 \\
    \hline

    \end{tabular}

    \label{tab: DT_final}
\end{table}

To facilitate user interaction, we implemented the decision tree in a web-based application with a graphical interface. The interface dynamically presents relevant questions and routes users through the decision logic based on their responses.

To assist users in interpreting legal terms and applying the classification logic, the tool includes several types of supplementary information: (1) guidance for understanding legal definitions, (2) decisions for example AI systems, (3) direct links to relevant AIA articles via the AI Act Explorer \cite{future_of_life_institute_eu_2025}, and (4) optional expert contact details. These support elements are embedded throughout the interface and can be accessed by clicking context-specific buttons. When clicked, a pop-up or expandable section reveals the relevant information directly within the user interface.

\subsection{Evaluation Design}
\label{subsec: Evaluation Design}

To assess the artifact, gather empirical insights, and thereby meeting requirement (ii) as stated in Section ~\ref{subsec: Artifact Design}, we conducted two evaluation phases using distinct experimental setups (Figure~\ref{fig: Research Process}): one qualitative and observed (EX\_beta), and one quantitative and unobserved (EX\_final).

\begin{itemize}
    \item[EX\_beta] Participants engaged with the artifact in the presence of at least two members of the evaluation team. Sessions were conducted remotely, with participants sharing their screens while applying the RCS. This setup enabled in-depth behavioral observation during the beta phase.
    \item[EX\_final] Participants used the artifact independently. User interaction data—such as time spent per question and use of support materials—was automatically logged. Additionally, participants completed a post-session Likert-scale questionnaire \cite{likert_technique_1932} to evaluate their experience with the artifact and provide subjective feedback on its usability and effectiveness.
\end{itemize}

The evaluation was conducted in two sequential phases: the qualitative phase (EX\_beta), followed by the quantitative phase (EX\_final). Both phases involved employees of a globally active company with over 100,000 employees operating across diverse sectors. Participants represented a wide range of roles, domains, and AI use cases. Importantly, the evaluation focused on the behavior of individual practitioners rather than organizational structures or centralized compliance units—reflecting also usage scenarios common in SMEs, decentralized teams, or public sector contexts.

\subsubsection{Artifact Beta Evaluation Phase}
\label{subsec: beta_evaluation}

The beta phase involved 11 practitioners, each applying the artifact to one of their real-world AI use cases using the EX\_beta setup. 

To ensure that our findings would be broadly relevant, we carefully selected AI use cases and participants to represent a wide range of domains, application types, and organizational roles. This purposive sampling aimed to capture diverse challenges and perspectives, thereby enhancing the generalizability of observations. A summary of participants, their domains, and expertise is provided in Table~\ref{tab: Overview_beta}.

During the sessions, evaluation team members took structured notes. After all sessions, recurring themes were synthesized, and key observations were clustered. To validate these qualitative findings, we formulated them into statements and returned them to participants for cross-validation using a five-point Likert scale (plus a “do not recall” option). This approach allowed findings from individual sessions to be cross-validated by all participants.

\begin{table}
\small
\caption{Overview of use cases analyzed in the beta evaluation phase. 
\newline
AIA Familiarity (self-assessed): NF = Not Familiar, SF = Slightly Familiar, MF = Moderately Familiar, VF = Very Familiar, E = Expert; 
\newline Expertise: B = Business, DIM = Data Science/Informatics/Mathematics, E = Engineering}
\label{tab: Overview_beta}
\centering
\begin{tabular}{|p{12 pt}|p{50 pt}|p{183 pt}|p{20 pt}|p{28 pt}|p{40 pt}|}
\hline
\textbf{ID} & \textbf{Domain} & \textbf{Short Description} & \textbf{AIA Fam.} & \textbf{Exp.} & \textbf{Role} \\
\hline
1 & Factory Automation & Generative AI-based automation engineering support. & VF & B, E, DIM & Innovation Manager \\
\hline
2 & Productivity Improvement & An application for identification and highlighting of legal clauses in a document, combined with guidance on how to amend the clauses. The application's goal is to save legal experts' time, reduce effort, and increase productivity. & VF & DIM & Head of AI Team \\
\hline
3 & Factory Automation & Automated adaptations of robotic cells to provide lot-size-one disassembly routines for EV batteries. & SF & E & Technical Developer \\
\hline
4 & Critical Infrastructure & AI application to get real-time insights into the state (voltage) of low voltage grids based on substation measurements. & MF & DIM & Technical Developer \\
\hline
5 & Comfort Optimization & AI application to fully automate and optimize the comfort (Heating, Ventilation, and Air Conditioning) operations in buildings. & SF & DIM & Project Manager \\
\hline
6 & Procurement & Application supporting the creation phase of service descriptions. The text input from the user is checked with the help of Generative AI for certain prerequisites, ensuring all relevant information can be found in the document. & SF & DIM & Technical Developer \\
\hline
7 & Rail Vehicles & Applies video analytics to live streams of CCTV cameras in rail vehicles (trains) and analyzes the live video streams for the occurrence of brawls, i.e., motion patterns that look like physical violence. & MF & E & Product Owner \\
\hline
8 & Productivity Improvement & Generative AI based chatbot providing users of the company-internal information distribution Web page for cybersecurity related topics easily with the right information. & VF & B & Cyber-security Professional \\
\hline
9 & Process Automation & A solution to automate request handling by classifying, information extraction, and further process triggering with NLP and GenAI technologies. & MF & B & Product Owner \\
\hline
10 & Factory Automation & A solution that assists operators and service technicians in resolving production line issues efficiently to maximize uptime. It provides a comprehensive 360-degree view of potential problems, including alarm codes, descriptions, sensor data, related manual information, and resolutions to similar issues. & MF & DIM  & Project Manager \\
\hline
11 & Productivity Improvement & A digital assistant for product- and tool-support. & MF & DIM & Product Owner \\
\hline
\end{tabular}
\end{table}

\subsubsection{Final Evaluation Phase}

The second evaluation phase involved 67 participants, recruited through internal company communication channels. Before using the artifact, all users completed a brief tutorial introducing the tool’s purpose, functionality, and available support materials. To ensure understanding, they were required to confirm completion via a checkbox before proceeding.

To ensure the reliability and relevance of the collected data, users were asked to provide contextual information about the AI system under evaluation, including system name, input/output modalities, intended users, and whether it was standalone or embedded. This metadata also supported the identification and filtering of non-serious or “test” entries.

Two data sets were collected: (1) user interaction logs (e.g., time per question, support feature usage, user-provided system descriptions), and (2) post-session questionnaire responses. These were linked for analysis and used for data credibility checks. Each participant was limited to a single submission.

\section{Evaluation Results}
\label{sec: Evaluation Results}
This section presents the findings from both the qualitative-oriented beta-evaluation phase and an analysis of the quantitative data obtained from the final evaluation phase.
\subsection{Qualitative results from the artifact beta evaluation phase}
\label{subsec: beta_eval}

Each paragraph first describes the findings we observed in the beta evaluation phase, followed by the direct implication this had for artifact design. Also, all findings are summarized in Table~\ref{tab:findings_implications}.
\begin{table}[ht]
\centering
\caption{Summary of findings and derived implications for artifact design.}
\begin{tabular}{|c|p{4cm}|p{7.5cm}|}
\hline
\textbf{No.} & \textbf{Finding} & \textbf{Implication for Artifact Design} \\
\hline
1 & Unclear terms and definitions in the RCS hinder consistent understanding & Provide additional clarifications and practical examples alongside legal definitions to support accurate user understanding. \\
\hline
2 & Accurate classification requires familiarity with EU harmonisation legislation & Include summaries, example products, and conditions requiring third-party conformity assessments for all regulations referenced in AIA Annex I. \\
\hline
 3& Clearly defined system purpose, functionality, and use context facilitates accurate classification & Request details on input, output, intended purpose, user group, and human oversight to help users articulate the system purpose, functionality, and use context and enable more reliable classification. \\
\hline
4 & The scope of the AIA itself should be considered & Prior to applying the RCS, implement a mechanism determining if a system is out of scope of the AIA to avoid unnecessary effort and guide user expectations. \\
\hline
5 & Access to multiple forms of support can aid users’ ability to self-classify AI systems with confidence & Incorporate layered support mechanisms throughout the tool, including contextual explanations, practical examples, references to legal text, expert contact options, and free-text fields. These features help users interpret questions, understand the rationale behind classification outcomes, and document their reasoning effectively. \\
\hline
6 & Question phrasing can lead to misinterpretation & Carefully phrase questions to balance ease of understanding for users with the need to preserve legal precision.  \\
\hline
\end{tabular}
\label{tab:findings_implications}
\end{table}
\subsubsection{Unclear terms and definitions in the RCS hinder consistent understanding}
The first finding from the sessions was that users tended to struggle with certain terms in the RCS and the corresponding definitions provided by the AIA.
Starting with the definition of "AI system" itself (Article 3 para. 1 AIA), users often found it difficult to determine if the system in question was covered by this definition. This supports the theoretical analysis in legal literature, criticizing parts of the definition of "AI systems" due to their lack of meaning and distinctiveness

\cite{wendehorst_begriff_2024}.
It is worth noting that in February 2025, after the completion of the data collection phase for this study, the European Commission published guidelines aimed at addressing ambiguities in the AIA’s definition of AI systems \cite{european_commission_commission_2025}. While this publication provides greater clarity, for example, by explicitly classifying machine learning (ML)-based approaches as AI, some uncertainties persist. Notably, the guidelines state that certain systems fall outside the scope of the definition "because of their limited capacity to analyse patterns and adjust autonomously their output" \cite{european_commission_commission_2025}. However, the term “limited capacity” remains undefined, leaving room for interpretation and continued uncertainty.

Another particularly challenging term was "safety component," which is crucial for the RCS as it is a key aspect in determining whether a system is classified as high risk.
Notably, the definition of "safety component" provided in the AIA differs from several other regulations mentioned in  AIA Annex I \cite{leinarte_classification_2024}, such as Regulation (EU) 1230/2023 (Machinery Regulation). 
The lack of congruence between definitions is challenging for users because a "safety component" defined by the AIA might not align with the definition in the Machinery Regulation, and vice versa.
This discrepancy can cause a practitioner familiar with another regulation's definition to misinterpret the term within the context of the AIA, potentially leading to misclassification without noticing during the classification process.
Moreover, even when the correct definition from the AIA is considered, its application to specific use cases remains challenging due to its vagueness and the absence of existing legal precedents.
Similarly, other terms such as "direct interaction with natural persons" and "synthetically generated content" led to uncertainties due to the lack of clear definitions in the AIA.

\textit{Derived artifact design implication:}
Providing legal definitions alongside interpretive guidance and concrete examples helps practitioners resolve ambiguities in regulatory text.
This combination supports more accurate application of the RCS, particularly for complex or loosely defined terms such as
“safety component.”

\subsubsection{Accurate classification requires detailed know-how on European Union harmonisation legislations}
According to AIA Article 6.1, one condition for classifying an AI system as high-risk is whether the system itself—or the product it is embedded in—falls under one of 20 Union harmonisation legislations (listed in Annex I of the AIA) and requires a third-party conformity assessment under that legislation. While this determination was straightforward for some use cases, it proved challenging in others—particularly when systems had multiple or evolving application contexts. Practitioners often struggled to judge whether their system might be subject to sector-specific regulation, leading to considerable uncertainty.

Importantly, this challenge extends beyond simply being aware of the 20 referenced legislations. Accurately determining whether a third-party conformity assessment is required demands in-depth knowledge of the specific scope, product definitions, and procedural requirements of each individual regulation. This level of legal and technical detail goes far beyond what most practitioners can reasonably be expected to know, making this requirement a significant barrier for any self-service implementation of the RCS.

\textit{Derived artifact design implication:}
To support accurate classification, tools implementing the RCS should include concise descriptions and illustrative examples for each referenced EU regulation. Where third-party conformity assessments are involved, integrated guidance should clarify when and why such assessments apply, helping practitioners bridge regulatory knowledge gaps without requiring deep legal expertise.

\subsubsection{Clearly defined system purpose, functionality, and use context facilitates accurate classification}
Users demonstrated greater confidence in answering classification questions when the purpose, functionality, and use context of the AI system was clearly defined. This is because the RCS is context-dependent, particularly for the unacceptable and high-risk categories, where the system’s intended purpose plays a central role in determining the appropriate classification. For example, determining whether an AI system qualifies as a safety component is essential for identifying it as high-risk. The more clearly the system's purpose, functionality, and use context is articulated, the easier it becomes to make such determinations. In contrast, users evaluating systems with a broad range of functionalities or systems still under development with unspecified features often struggled to classify them confidently.

\textit{Derived artifact design implication:}
Prompting users to specify an AI system’s input/output modalities, intended purpose, user group, and oversight conditions improves classification accuracy. While this information may not influence decision logic directly, it helps users frame the system's purpose, functionality, and use context more clearly and apply the RCS with greater confidence.

\subsubsection{The scope of the AIA itslef should be considered}
We also noted the significance of determining whether a system under consideration falls within the scope of the AIA. Article 2 outlines various conditions that dictate the applicability of the AIA to a specific AI system, such as geographical scope and exceptions for certain types of AI systems. Participants found this information particularly valuable, as it could influence future product development.
Furthermore, certain areas of product legislation are out of scope (AIA Article 2 para. 3 and Annex I Section B).
However, due to amendments to the aforementioned legislation, the AIA may have only indirect effects beyond the scope of the RCS. Albeit an AI system may be qualified out of scope for a certain intended use, the RCS includes further risk assessment if the intended use of the AI systems exceeds the scope of Article 2 para. 3 AIA or has multiple intended use cases (e.g. AI-based assistance systems in motor vehicles and other machines).

\textit{Derived artifact design implication:}
Before applying the RCS, there should be a mechanism to determine whether an AI system falls outside the scope of the AI Act, as this can streamline the classification process and inform product development decisions.

\subsubsection{ Access to multiple forms of support can aid users’ ability to self-classify AI systems with confidence}
The previous paragraphs already outline the value of supplementary information for particular aspects.
We additionally found that the provision of such information improves the transparency of the RCS, increasing participants' confidence. 
While references to the original legal text and transparent explanations of the questions' rationales and the consequences of the answers (e.g., "this question determines if your AI system is classified as an unacceptable risk") were perceived as supportive, participants found the most value in support mechanisms such as explanations for unclear terms and definitions, as well as practical examples. Additionally, providing the contact information of an available expert for addressing uncertainties was evaluated as particularly beneficial. This finding is further examined in the final phase’s quantitative analysis. In addition, users appreciated the option to justify their responses in free-text fields—especially when facing uncertainty. While this input did not influence the classification outcome, it enhanced transparency, documentation, and user acceptance.

\textit{Derived artifact design implication:} The artifact should incorporate multiple layers of support: contextual definitions, example-based guidance, links to legislative sources, access to expert contact, and free-text fields to document user rationale.

\subsubsection{Question phrasing can lead to misinterpretation}

Question wording had a notable effect on how participants interpreted and answered items in the RCS. For example, when identifying prohibited AI practices (Art. 5 AIA), several participants considered hypothetical misuse scenarios rather than the intended use. This often led to misclassification and stemmed from Article 5’s legal framing, which emphasizes actions and outcomes, typical of tort law, rather than system-level characteristics, as seen in other parts of the AIA (e.g., Art. 6).

\textit{Derived artifact design implication:}
When translating the RCS into a questionnaire, designers must carefully translate regulatory language to reflect the legislator’s intent while preserving legal accuracy and ensuring accessibility for non-experts. Achieving this requires balancing legal precision, conceptual clarity, and practical understandability.

\subsubsection{Participant cross-check}
As described in Section~\ref{sec: Methodology}, to further investigate the effects of the different support mechanisms, based on the results from the sessions, we derived statements, which were formulated into a 5-point Likert scale questionnaire, with an additional option “I do not remember.”  Ten of the eleven participants responded.

To summarize the results, we used the Interpolated Median (IM) \cite{lorenz_ranking_2018} and Percent Favourable (PF) \cite{nulty_adequacy_2008}, both of which are appropriate for balanced Likert data \cite{zumrawi_proposed_2023}. The IM accounts for skew around the median, while PF measures the proportion of positive agreement (“agree” or “strongly agree”).

\begin{figure}[ht]
    \centering
    \includegraphics[width=1\linewidth]{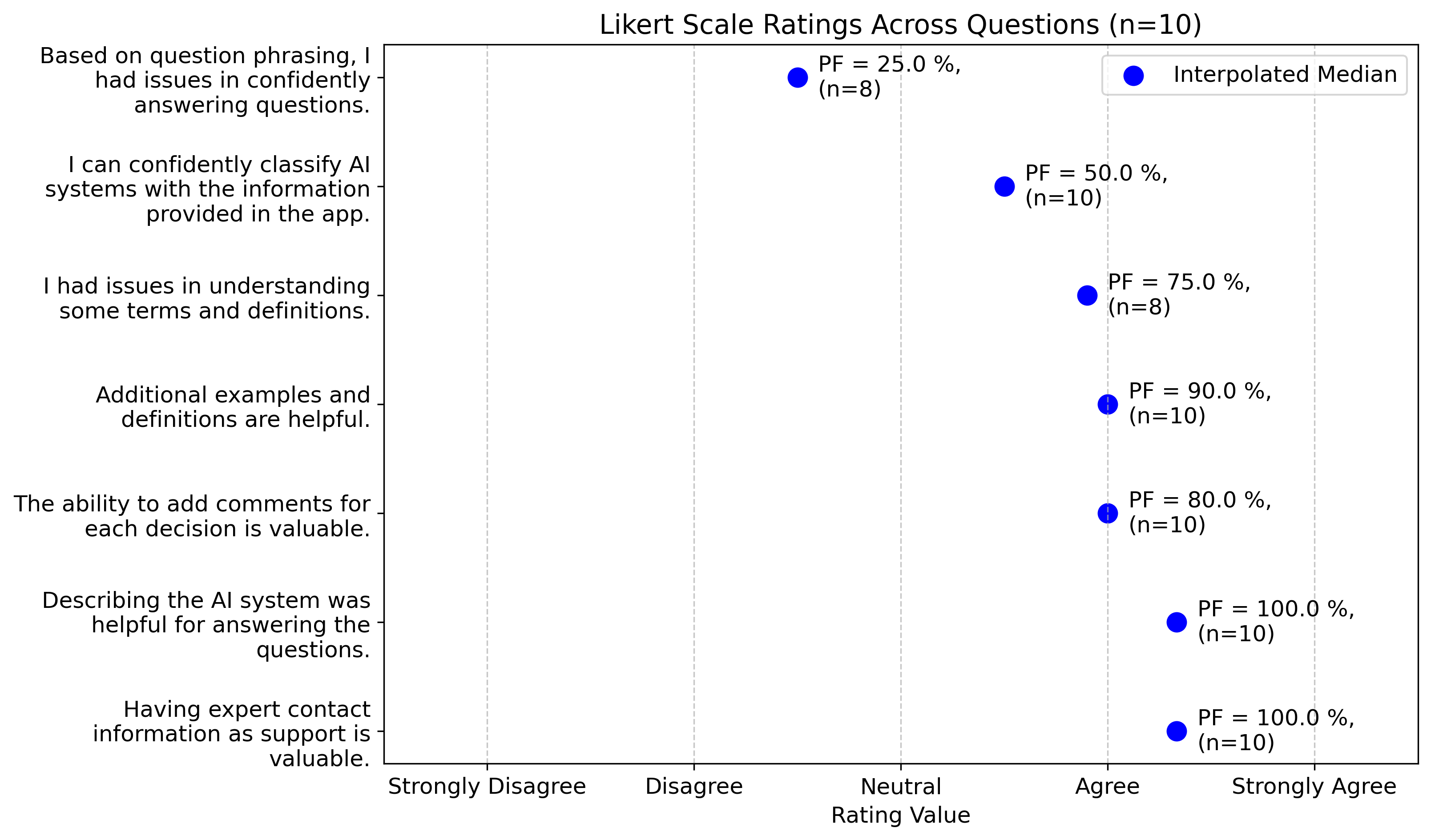}
    \caption{Interpolated Median \cite{lorenz_ranking_2018} and Percent Favourable \cite{nulty_adequacy_2008} evaluated for all statements in the participant cross-evaluation (Ex\_beta).}
    \label{fig: Likert_beta}
\end{figure}

The results of this analysis are shown in Figure~\ref{fig: Likert_beta}.

A notable outcome was the statement regarding difficulties in answering classification questions due to their wording: only 25.0\% of participants agreed. This low agreement may be explained by the evaluation setup: during the beta phase, evaluation team members were present and, in some cases, provided clarifications when participants explicitly asked. This occasional direct assistance may have reduced the likelihood of reported difficulties related to question formulation.
Another key insight is that half of the participants agreed that they felt confident applying the RCS using the tool as a self-service support (PF = 50.0\%). The remaining items received overall agreement.

\subsection{Quantitative results from the second evaluation phase}
The aim of the second evaluation phase was to confirm or refute the findings from the artifact beta evaluation phase, utilizing an experimental setup that avoids direct observation by the evaluation team.

The first main finding we sought to validate was the importance of different types of supplementary materials. To achieve this, we analyzed tracking data from all 67 users included in this study. The tracking data captured the number of times users clicked on a button to access various supplementary materials for each user journey. Figure~\ref{fig: Buttons} illustrates the frequency distribution of accesses, categorized by the type of supplementary material.
\begin{figure}[ht!]
\centering \includegraphics[width=0.8\linewidth]{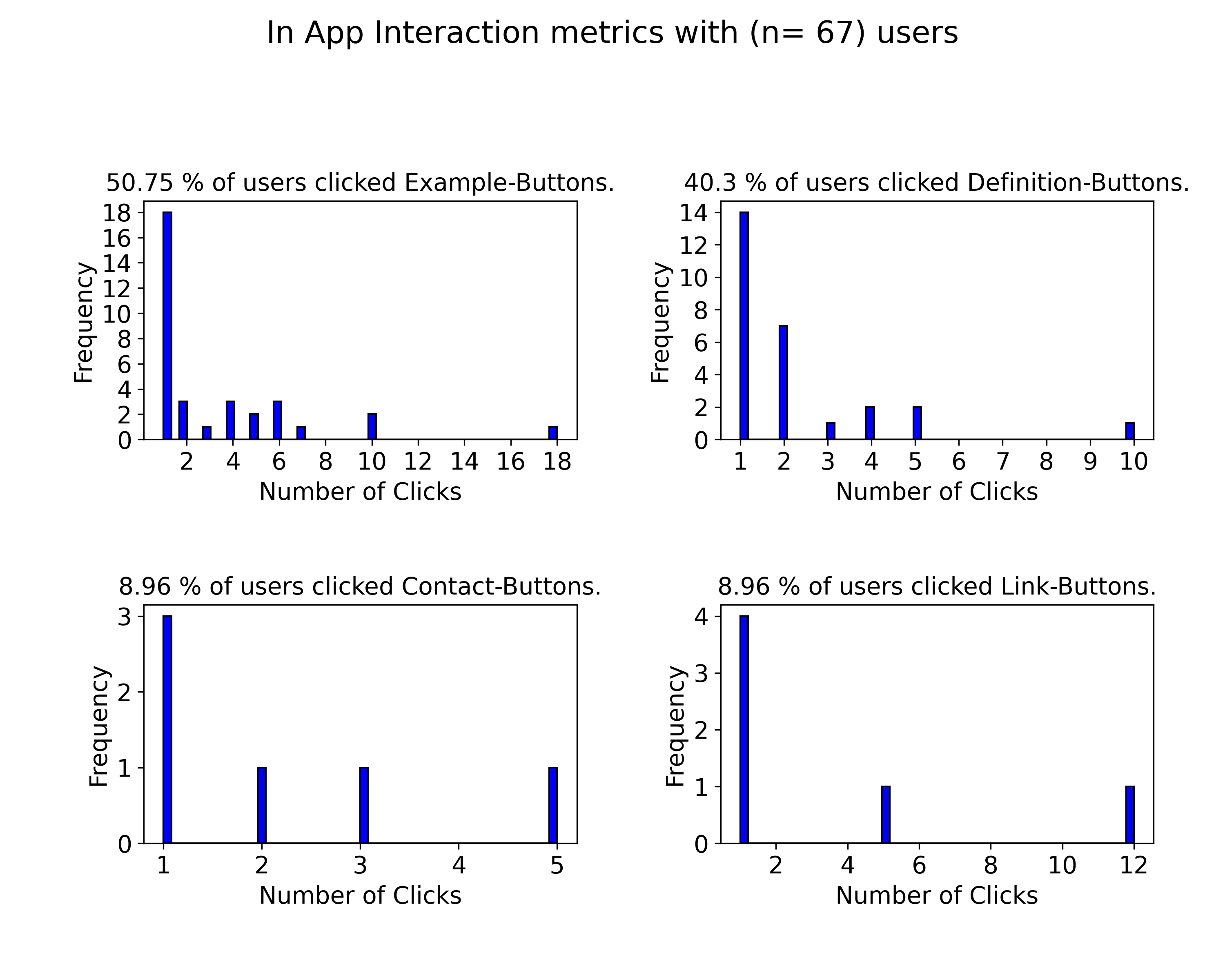} \caption{Distribution of accesses to supplementary materials per user, grouped by types of supplementary material. The title of each subplot denotes the percentage of users that accessed the respective supporting material at least once throughout their use.} \label{fig: Buttons} \end{figure}

Based on these distributions, we verified that definitions and examples are the most sought-after types of supplementary material. Although the beta evaluation phase highlighted the significance of having access to a (legal) expert, this was not reflected in the number of button accesses for this material type, as only 8.96\% of users actively selected it. 

\begin{figure}[ht!]
    \centering
    \includegraphics[width=0.9\linewidth]{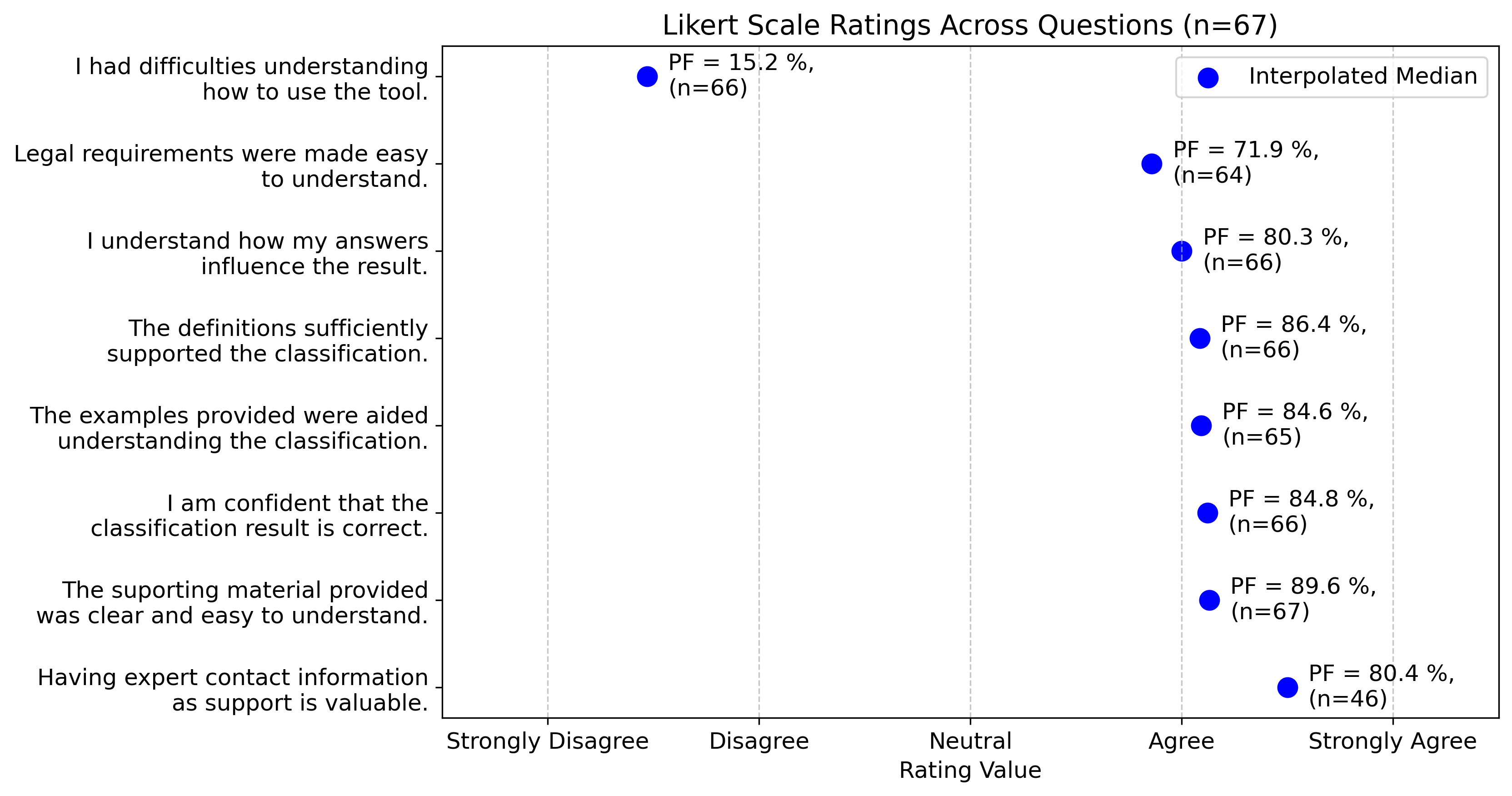}
    \caption{Interpolated Median \cite{lorenz_ranking_2018} and Percent Favourable \cite{nulty_adequacy_2008} evaluated for all statements in the final evaluation Likert-Scale survey (Ex\_final).}
    \label{fig: Likert}

\end{figure}

Nevertheless, the Likert-scale-based user questionnaire indicates that the availability of an expert is highly valued, receiving the highest score on the Likert scale in terms of the IM, as shown in Figure~\ref{fig: Likert}. As in the beta phase, we analyzed Likert-scale responses using the IM and PF metrics.
Overall, the results indicate strong agreement across all statements, supporting the effectiveness of the implemented support mechanisms within the artifact. Furthermore, the statement regarding the helpfulness of having an expert contact person received the highest level of agreement, as reflected by the IM score.
Although this item received the highest IM score, its PF score was comparatively lower, indicating a more polarized response distribution.
These findings align with our observations from the artifact beta evaluation phase: when an expert is needed, their presence is considered very helpful. However, when an expert is not needed, participants tended to respond neutrally or express disagreement.

To identify specific problem areas, we asked participants whether they encountered difficulties with particular questions. 21 out of 67 participants agreed that they did.
We offered the opportunity for participants to provide explanations, and 9 specifically mentioned difficulties with the question addressing the potential coverage by other Union harmonisation legislation and the corresponding required third-party conformity assessment.
This supports the finding from the beta evaluation phase that this is one of the main challenges in the RCS.

\section{Limitations and Discussion}
\label{sec: Discussion and Limitations}

This section details the limitations encountered during the study and respective mitigation strategies employed. Also, we provide an overview of the answers to the research questions scoped in Section~\ref{sec: Introduction} and derive implications for legislators and practitioners.
\subsection{Study limitations}

Firstly, the study focused on practitioners within an industrial setting. Consequently, the targeted participant group was composed of employees from a globally acting company with more than 100,000 employees spanning multiple sectors. To ensure the representativeness of the broader population, a careful selection of use cases and participant profiles was conducted (Section~\ref{subsec: beta_evaluation}). Furthermore, our experimental setup observed the decision-making of individuals without utilizing the expertise available in other parts of the company. This setup approximates conditions in SMEs or public institutions, where such expertise is often unavailable. Nevertheless, we encourage other scholars to extend this research into this problem setting within other demographics to build upon our findings.

Secondly, this study employed two distinct experimental setups, each with its own known strengths and limitations. In experimental setup EX\_beta, participants were actively observed while applying the RCS, which allowed the evaluation team to gather detailed qualitative insights into participant behavior. However, the presence of the evaluation team poses the risk of introducing observation biases.
Conversely, experimental setup EX\_final relied on indirect observation, utilizing in-app tracking data and responses from a Likert-scale-based questionnaire. This setup minimized observation bias since no evaluation team members were present during data collection. Also, the data gathered from EX\_final created empirical evidence to validate the findings derived from the qualitative data obtained from EX\_beta.
In this study, we adopted a mixed-methods approach to capitalize on the benefits of both methodologies, facilitating a cross-method evaluation to enhance the robustness of our findings.

Lastly, this study and the RCS within is strictly limited to AI systems regulated by the EU AIA, excluding in particular AI models and other AI-related legislation, e.g. AI systems under Annex I Section B, AI used in products subject to the Machinery Regulation and/or subject to the Regulation (EU) 988/2023 (General Product Safety Regulation).
Additionally, the RCS implemented in this study does not cover the risk classification scheme for General Purpose AI (GPAI) models, which is also defined in the AIA. 
However, the classification scheme for GPAI models is independent of the RCS for AI systems addressed in this study.
A detailed analysis of the classification scheme for GPAI is reserved for future work.

\subsection{Research Insights and Implications for Regulatory Authorities and Practitioners}
This subsection summarizes the insights gained in response to the three research questions introduced in Section~\ref{sec: Introduction}.
\paragraph{RQ1: What level of expertise is required to effectively apply the AI Act Risk Classification Scheme?}
The results show that accurately applying the RCS requires a combination of legal, technical, and domain-specific expertise, along with a working understanding of various EU harmonization laws.
Legal expertise is necessary to interpret key regulatory terms and definitions in the AIA. Technical and domain knowledge is essential to assess the system’s functionality, intended purpose, and potential impact.
In particular, familiarity with the scope and application of EU legislation referenced in Annex I is required to determine whether third-party conformity assessments apply—a key condition for high-risk classification under Article 6.
Although straightforward classification was possible in certain cases, it was often complicated by system complexity, evolving use cases, or regulatory ambiguity.
Since the required level of expertise cannot be assumed in all settings—especially in SMEs or early-stage development contexts—structured guidance and support mechanisms are essential for enabling a self-service risk classification tool. These may come from legislators directly or be embedded into third-party tools. The specific forms of effective support are discussed further under RQ3.

\paragraph{RQ2: What are the key challenges practitioners face in the AI Act Risk Classification Scheme, and how can these be addressed?}
The main challenges for practitioners in conducting the RCS, as identified in this study, are:

\begin{itemize}
    \item \textbf{Ambiguity in legal definitions:} Various legal terms in the AIA, such as "safety component," "direct interaction with natural persons," and "use of synthetically generated content," lack precise definitions, leading to difficulties in interpretation and application.

    \item \textbf{Regulatory overlap and inconsistencies:} AI systems often fall within multiple regulatory frameworks (e.g., the Machinery Regulation), which may contain conflicting definitions or requirements, making classification more complex.
    
    \item \textbf{Challenges in defining system scope and purpose:} Accurately determining an AI system’s risk class requires a clear understanding of its intended purpose, functionality, and deployment context. However, when systems are still in development or designed for multiple use cases, precisely defining what constitutes “the system” for classification purposes can be difficult, leading to potential misclassification.
    \item \textbf{Misinterpretation and uncertainty:} We observed several cases where practitioners misinterpreted key terms and aspects of the RCS, leading to misclassification. Additionally, when practitioners were uncertain about a specific part of the RCS, they may proceed with a potentially incorrect answer, further increasing the risk of misclassification.

\end{itemize}
Together, these challenges reveal that the RCS, while conceptually structured, presents significant interpretive hurdles in practice, particularly for non-experts.
\paragraph{RQ3: What specific information or resources would enhance practitioners’ ability to apply the AI Act Risk Classification Scheme effectively?}
The experimental results indicate that practitioners would benefit most from access to experts in the fields identified in the response to RQ1, as this was mentioned as most demanded support mechanism in the user survey. However, if such expertise is unavailable or too expensive, alternative support mechanisms can enhance the application of the RCS.

First, additional guidance, including comprehensive and understandable interpretations of the aforementioned legal definitions and terms, would significantly improve practitioners’ ability to classify AI systems correctly. While the European Commission has published some guidance on interpreting the definition of AI systems \cite{european_commission_commission_2025}, this represents only a first step. Our analysis identifies many additional aspects within the RCS that require further clarification.

Second, providing examples of correctly classified AI systems with detailed rationales for each classification step is highly beneficial, as this was the most frequently accessed support mechanism in our experiments. We infer that an official catalog of classified examples would be highly recommended, enabling practitioners to better understand and apply the RCS to their specific use cases, thereby improving classification accuracy and consistency.

Providing references to the actual wording of the legislator for relevant parts in the RCS was the least effective support mechanism in this study, suggesting that raw legal text alone may be insufficient for practical application.

\section{Conclusion}
\label{sec: Conclusion}

This study explored the practical application of the RCS by designing and evaluating a web-based decision-support tool. Through a two-phase empirical evaluation, we gathered qualitative and quantitative insights into how industrial practitioners engage with the RCS in real-world settings.

Our findings reveal that accurate application of the RCS requires a combination of legal expertise, knowledge of EU technical regulations and standards, and a thorough understanding of the AI system in question. Key challenges stem from ambiguous definitions in the AIA and the need to interpret a wide range of harmonized EU legislation. These issues pose substantial barriers to apply the RCS, especially in self-service settings or for users without legal or regulatory expertise.
To address these challenges, we implemented support mechanisms such as contextualized definitions, practical examples of classified AI systems, and access to expert guidance. These mechanisms were shown to improve user confidence and support more consistent classification outcomes.

These findings underscore the need for user-centered compliance tools that make regulatory requirements actionable for practitioners without legal training.

\bibliographystyle{splncs04}
\bibliography{references}

\end{document}